\documentstyle[seceq,preprint,epsf]{ptptex}

\def\su{\ifmmode{\tilde{u}} \else{$\tilde{u}$} \fi}
\def\sd{\ifmmode{\tilde{d}} \else{$\tilde{d}$} \fi}
\def\sq{\ifmmode{\tilde{q}} \else{$\tilde{q}$} \fi}
\def\sg{\ifmmode{\tilde{g}} \else{$\tilde{g}$} \fi}
\def\snu{\ifmmode{\tilde{\nu}} \else{$\tilde{\nu}$} \fi}
\def\se{\ifmmode{\tilde{e}} \else{$\tilde{e}$} \fi}
\def\smu{\ifmmode{\tilde{\mu}} \else{$\tilde{\mu}$} \fi}
\def\sfp{\ifmmode{\tilde{f}_{1L}} \else{$\tilde{f}_{1L}$} \fi}
\def\sfm{\ifmmode{\tilde{f}_{2L}} \else{$\tilde{f}_{2L}$} \fi}
\def\simgt{\rlap{\lower 3.5 pt \hbox{$\mathchar \sim$}}%
           \raise 1pt \hbox {$>$}}
\def\simlt{\rlap{\lower 3.5 pt \hbox{$\mathchar \sim$}}%
           \raise 1pt \hbox {$<$}}

\notypesetlogo  

\preprintnumber{
KEK-TH-459\\December 1995\\hep-ph/9512227}

\title{%
Electroweak Precision Measurements and the Minimal Supersymmetric
Standard Model%
\footnote{Talk given by Y.~Yamada
at Yukawa International Seminar (YKIS) '95
``From the Standard Model to Grand Unified Theories'',
Kyoto, Japan, August 21--25, 1995. To appear in the Proceedings.
}
}

\author{%
Youichi Yamada\footnote{Present Address; Physics Department,
University of Wisconsin, Madison, WI 53706, USA},
Kaoru Hagiwara and Seiji Matsumoto
}

\inst{%
Theory Group, KEK, 1-1 Oho, Tsukuba, Ibaraki 305, Japan
}

\recdate{%
\today
}

\abst{%
We discuss supersymmetric contributions to the electroweak precision
measurements in the Minimal Supersymmetric Standard Model for two
cases: the quark-lepton universality violation $\delta_{q\ell}$
in charged currents
and the ratio $R_b=\Gamma(Z\rightarrow b\bar{b})/\Gamma(Z\rightarrow
\rm hadrons)$. The recent experimental data suggest deviations from the
Standard Model for these observables, at the 1-$\sigma$ level for the former
and at more than 3-$\sigma$ level for the latter.
We analyze the non-oblique corrections from the SUSY particles to
explain these discrepancies.
The observed non-zero $\delta_{q\ell}$ may be explained by relatively light
sleptons, charginos and neutralinos. Although the observed excess of $R_b$
can be explained by very light scalar top and chargino for
low $\tan\beta$, this interpretation is
severely constrained by the opening of exotic decay modes for the
top quark.
}

\begin{document}

\maketitle

\section{Introduction}

Among many possible extensions of the Standard Model (SM), the Minimal
Supersymmetric (SUSY) Standard Model (MSSM) \cite{mssm} has attracted
much interest because it stabilizes the hierarchy between the grand
unification scale and the electroweak scale, and at the same time it gives
grand unification of the three gauge couplings consistent with the
precision measurements.
However, to solve the hierarchy problem, the MSSM has to contain
many new particles, SUSY particles, near the present energy frontier.
It is therefore expected that properties of these new particles may be
probed by the present precise measurements of the electroweak processes,
both at low energies and at the $Z$ boson pole.

In general, loop corrections of the SUSY particles to
the electroweak four-fermi processes
are classified into two categories: the corrections to the vector boson
propagators (oblique corrections) and the vertex and box corrections
(non-oblique corrections). The oblique corrections are universal for various
processes and have been extensively studied. In contrast, the non-oblique
corrections are highly process-dependent. One therefore has to specify the
process for the detailed study of these corrections.

In this paper we discuss two examples of the precision measurements
where the non-oblique SUSY corrections play a crucial role.
First, we study violation of the quark-lepton
universality in charged current interactions. Second, we study
deviation of the ratio $R_b=\Gamma(Z\rightarrow b\bar{b})%
/\Gamma(Z\rightarrow\rm hadrons)$ from the SM prediction,
the long-standing problem in the LEP/SLC precision measurements.
These examples are also interesting because the present experimental data
suggest the discrepancy with the SM, at the 1-$\sigma$ level for the former
and at more than 3-$\sigma$ level for the latter.

This paper is organized as follows.
In section 2, we discuss violation of the quark-lepton universality
in charged currents. In section 3, we discuss a possible solution of the
$R_b$ problem in the MSSM. Section 4 is devoted for summary.

\section{Quark-lepton universality violation in
charger currents}

The tree-level universality of the charged current weak interactions is one
of the important consequences of the SU(2)$_L$ gauge symmetry of the
fundamental theory.
The universality between quarks and leptons is expressed
as the unitarity of the Cabibbo-Kobayashi-Maskawa (CKM) matrix, for example,
\begin{equation}
|V_{ud}|^2+|V_{us}|^2+|V_{ub}|^2=1. \label{1}
\end{equation}
Present experimental data\cite{towner,sirlin2,rpp}
for these CKM matrix elements are
extracted from the ratios of the amplitude of the
semileptonic hadron decays to that of the muon decay, after subtracting
the known SM radiative corrections \cite{sirlin,leut}.
Here we adopt the following values
\begin{equation}
|V_{ud}|=0.9745\pm0.0007[\citen{towner},\citen{sirlin2}],\;\;
|V_{us}|=0.2205\pm0.0018[\citen{rpp}],\;\;
|V_{ub}|=0.003\pm0.001[\citen{rpp}]. \label{2}
\end{equation}
Their squared sum is then
\begin{equation}
|V_{ud}|^2+|V_{us}|^2+|V_{ub}|^2-1=-0.0017\pm0.0015. \label{3}
\end{equation}
The universality is violated at the 1-$\sigma$ level.

In general, the universality (\ref{1}) can be modified by
process-dependent radiative corrections, due to the spontaneous
breaking of the SU(2)$_L$ symmetry.
The data (\ref{2}) lead to a small deviation from
the quark-lepton universality (\ref{3}), after the SM corrections are
applied.
This may suggest a signal for physics beyond the SM, although it is only
at the 1-$\sigma$ level.

Here we study the SUSY one-loop contribution to the
quark-lepton universality violation in the low-energy charged
currents \cite{hmy}.
We refine previous works \cite{barbieri1,barbieri2} by extending
the analysis to cover the whole parameter space of the MSSM
and obtain constraints on SUSY particle masses from the 1-$\sigma$
universality violation (\ref{3}).

We study the decay $f_2\to f_1\ell^-\bar{\nu}$
where $f=(f_1, f_2)$ is a SU(2)$_L$ fermion doublet,
comparing the following two cases;
muon decay $(f_1=\nu_{\mu}, f_2=\mu^-)$ and
semileptonic hadron decays $(f_1=u, f_2=d, s, b)$.
At the tree level, their decay amplitudes are identical
up to the CKM matrix element, and are
expressed in terms of the bare Fermi constant $G_0=g^2/4\sqrt{2}m_W^2$.
At the one-loop level, however, the decay amplitudes
receive process-dependent non-oblique corrections as well
as oblique ones.
Following the formalism in Ref.\citen{hhkm}, the corrected decay amplitudes
of $f_2\to f_1\ell^-\bar{\nu}$ are expressed as
\begin{equation}
G_{f}=\frac{\bar{g}_W^2(0)+g^2\bar{\delta}_{Gf}}
{4\sqrt{2}m_W^2}. \label{4}
\end{equation}
The effective coupling $\bar{g}_W^2(0)$ contains the correction to the
W-boson propagator and does not lead to the universality violation.
The process-dependent term $\bar{\delta}_{Gf}$, which represents the
vertex and box corrections, gives the universality violation.

In the MSSM,
$\bar{\delta}_{Gf}=\bar{\delta}_{Gf}({\rm SM})+\delta_{Gf}({\rm SUSY})$,
where $\bar{\delta}_{Gf}({\rm SM})$ is the gauge vector
loop contribution and $\delta_{Gf}({\rm SUSY})$ is the SUSY loop
contribution. Since $\bar{\delta}_{Gf}({\rm SM})$ has been subtracted in
extracting the data (\ref{2}), it is $\delta_{Gf}({\rm SUSY})$ which gives
the universality violation shown in (\ref{3}).
$\delta_{Gf}({\rm SUSY})$ comes from the loops with left-handed
sfermions ($\snu_e$, $\se$, $\tilde{f}_1$, $\tilde{f}_2$)$_L$,
charginos $\tilde{C}_j(j=1,2)$, neutralinos $\tilde{N}_i(i=1-4)$
and a gluino \sg.
It is expressed as a sum
$\delta_f=\delta_f^{(v)}+\delta_{\ell}^{(v)}+\delta_f^{(b)}$, where we use the
abbreviations $\delta_f\equiv\delta_{Gf}({\rm SUSY})$ etc.
The explicit forms of the correction $\delta_f^{(v)}$ to
the $W^+\bar{f}_1f_2$ vertex and the box correction $\delta_f^{(b)}$ are
given in Ref.\citen{hmy}. They are functions of the masses and mixing matrices
of the above SUSY particles.

The quark-lepton universality violation (\ref{3}) is now expressed as
\begin{eqnarray}
\delta_{q\ell}\equiv
\frac{\delta G_q}{G_q}-\frac{\delta G_{\mu}}{G_{\mu}}&\equiv &
(|V_{ud}|^2+|V_{us}|^2+|V_{ub}|^2)^{\frac{1}{2}}-1 \nonumber \\
&=&\delta_q-\delta_{\ell} \nonumber \\
&=&(\delta_q^{(v)}+\delta_{\ell}^{(v)}+\delta_q^{(b)})
-(2\delta_{\ell}^{(v)}+\delta_{\ell}^{(b)}) \nonumber \\
&=&(\delta_q^{(v)}+\delta_q^{(b)})
-(\delta_{\ell}^{(v)}+\delta_{\ell}^{(b)}) \nonumber \\
&=&-0.0009\pm 0.0008. \label{8}
\end{eqnarray}

We now give numerical estimates of $\delta_{q\ell}$.
We assume generation independence of the sfermion masses,
and impose the following mass relations suggested by the minimal
supergravity model with grand unification \cite{nilles}
\begin{equation}
M_1=\frac{5}{3}M_2\tan^2\theta_W,\;\;\;
M_{\sg}=\frac{g_s^2}{g^2}M_2,\;\;\;
M_{\tilde{Q}}^2=M_{\tilde{L}}^2+9M_2^2,
\label{9}
\end{equation}
to reduce the number of independent parameters 

In Fig.1, the quark-lepton universality violation
$\delta_{q\ell}$ is shown in the ($M_2$, $\mu$) plane for $\tan\beta=2$ and
$m_{\snu}=$(50, 100)~GeV. The solid lines are contours for
constant $\delta_{q\ell}$'s.
The regions below the thick solid lines ($\delta_{q\ell}=-0.0001$)
are consistent with the 1-$\sigma$
universality violation (\ref{3}). The regions below the thick dashed lines
are excluded by LEP-I experiments because SUSY particles have not been
observed in $Z$ decays. Therefore the regions under the
thick solid lines and above the thick dashed lines are favored
by the present data.
As seen in the figure, the SUSY parameters which
satisfy the universality violation (\ref{3}) and the LEP-I bound tend to lie
in the $M_2\simlt|\mu|$ region, where the lighter chargino and neutralinos are
gaugino-like. On the other hand, when
$M_2\simgt|\mu|$, $\delta_{q\ell}$ tends to be positive and disfavors
the negative deviation (\ref{3}). We also find that the $\tan\beta$
dependence is not significant and that the gluino contribution
to $\delta_{q\ell}$ is completely negligible, less
than ${\cal O}(10^{-6})$ in magnitude.
\begin{figure}
\begin{minipage}[t]{75mm}
\epsfxsize=70mm
\centerline{\epsffile{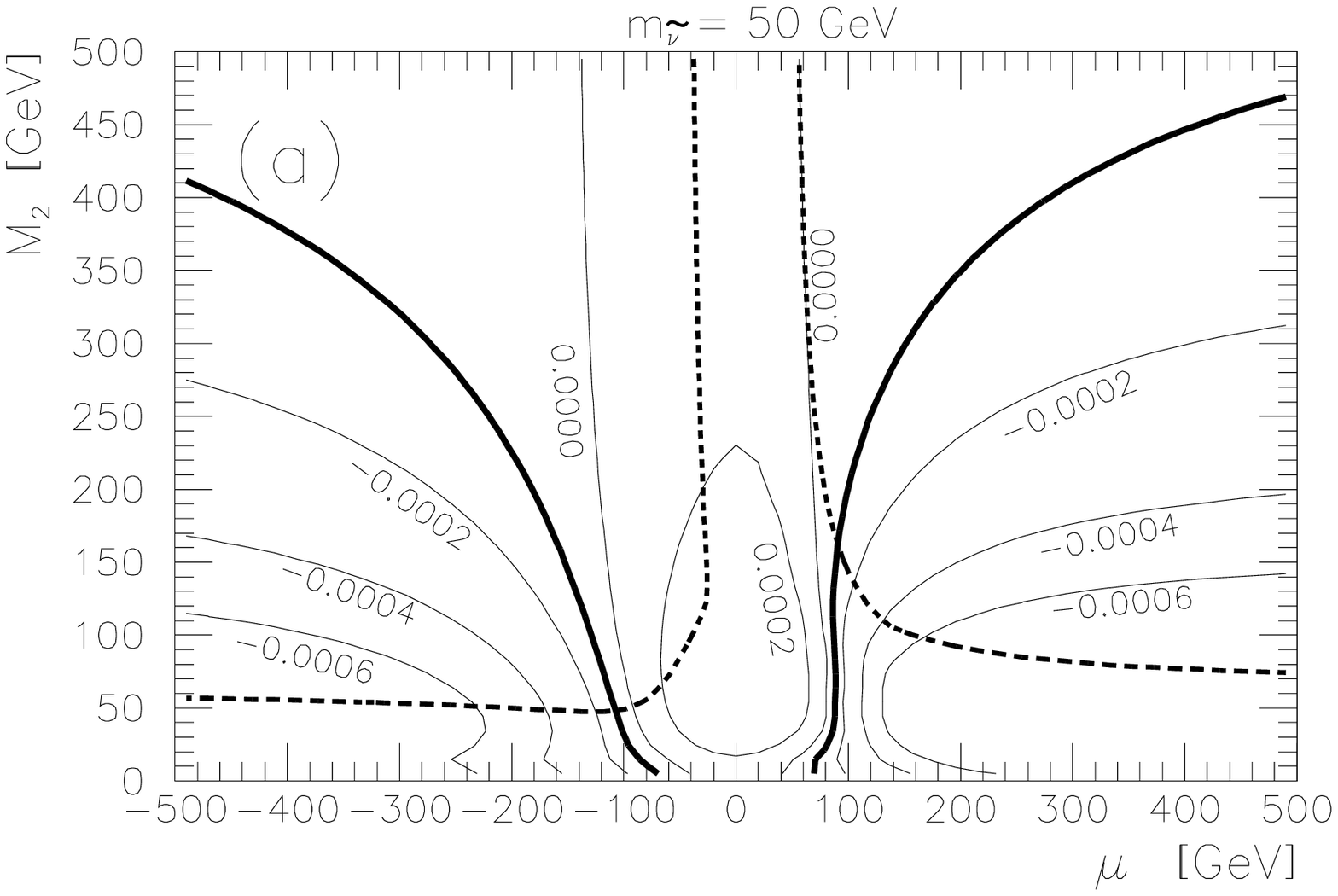}}
\end{minipage}
\begin{minipage}[t]{75mm}
\epsfxsize=70mm
\centerline{\epsffile{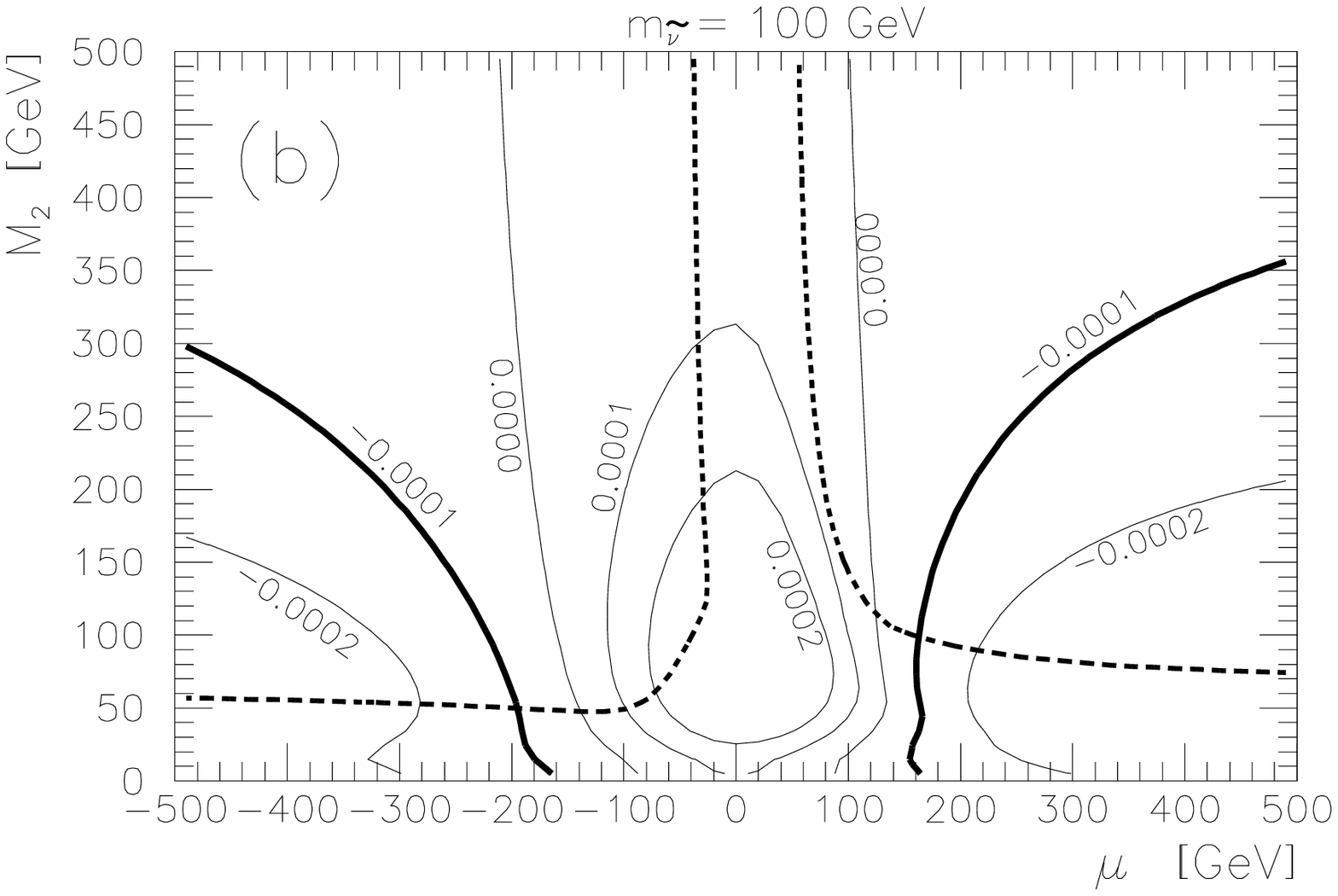}}
\end{minipage}
\caption{
The SUSY contribution to the quark-lepton universality violation
parameter $\delta_{q\ell}$ in the $(M_2, \mu)$ plane for
$\tan\beta=2$ and $m_{\snu}({\rm GeV})=$50(a), 100(b).
The SUSY contribution explains the universality violation (2$\cdot$3)
at the 1-$\sigma$ level in regions below the thick solid lines.
The regions below the thick dashed lines are excluded by LEP-I
experiments.
}
\end{figure}

In Fig.2, the 1-$\sigma$ allowed region of masses of the
sneutrino \snu and the lighter chargino $\tilde{C}_1$ from
the quark-lepton universality violation (\ref{3}) is shown.
The 1-$\sigma$ (67\% C.L.) upper bounds are roughly
$m_{\snu}<220$~GeV and $m_{\tilde{C}_1}<600$~GeV, respectively.
Therefore, the 1-$\sigma$ deviation (\ref{3}) from the
quark-lepton universality tends  to favor light sleptons
and relatively light chargino and neutralinos with significant
gaugino components.
It is interesting that while the upper bound of $m_{\snu}$ generally
decreases with increasing $m_{\tilde{C}_1}$, it increases between
$m_{\tilde{C}_1}\simeq 50$~GeV and $m_{\tilde{C}_1}\simeq 100\;{\rm GeV}$,
similar to the case of $\delta_{\ell}$ that has been
studied in Ref.\citen{chankow}.

\begin{figure}
\epsfxsize=8cm
\centerline{\epsffile{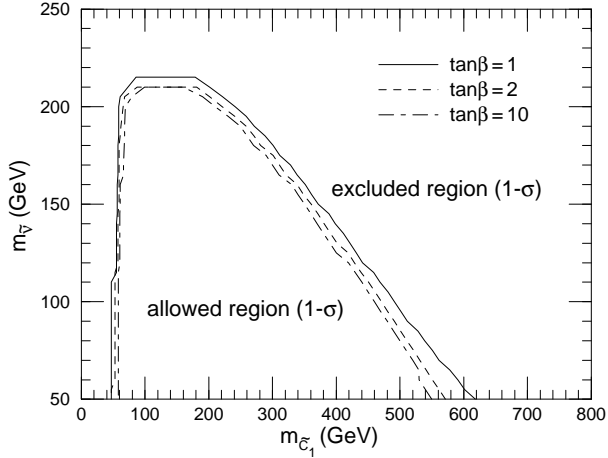}}
\caption{
The 1-$\sigma$ allowed region of ($m_{\tilde{C}_1}$, $m_{\snu}$)
for explaining the universality violation (2$\cdot$3) as the SUSY
contribution for $\tan\beta=(1,2,10)$.
}
\end{figure}

Finally, we comment on the sign of $\delta_{q\ell}$.
As seen in Fig.1, $\delta_{q\ell}$ takes both signs,
contrary to the observation of Ref.\citen{barbieri1} where only cases with
very light sfermions ($M_{\tilde{L}}<m_Z/2$, $M_{\tilde{Q}}<m_Z$)
were studied and
only negative $\delta_{q\ell}$ was found.
This sign change is caused by the cancellation\cite{hmy} between
the vertex and the box corrections in (\ref{8}).

\section{$R_b$ problem}

\setcounter{equation}{0}
The present precision measurements of the electroweak processes at the Z pole
show an excellent agreement with the Standard Model for the observed top quark
mass,
\begin{equation}
m_t({\rm GeV})= \left\{ \begin{array}{l} 176\pm 8\pm 10 [\citen{mt1}], \\
199\pm{}^{19}_{21}
\pm 22 [\citen{mt2}],
\end{array} \right.
\end{equation}
except for the ratios
$R_b=\Gamma(Z\rightarrow b\bar{b})/\Gamma(Z\rightarrow\rm hadrons)$
and
$R_c=\Gamma(Z\rightarrow c\bar{c})/\Gamma(Z\rightarrow\rm hadrons)$.
The discrepancy in
these ratios, sometimes called ``$R_b$ crises'', is a long-standing problem
in the precision test of the SM.
The most recent preliminary data\cite{lep95,hagiwara} for the 1995
summer conferences show
\begin{equation}
R_b(1995)=0.2219\pm0.0017,\;\;
R_c(1995)=0.1543\pm0.0074, \label{eq31}
\end{equation}
for the fit with unconstrained $R_c$ and
\begin{equation}
R_b(1995)=0.2206\pm0.0016, \label{eq32}
\end{equation}
for the fit with fixed $R_c=0.171$ (the SM prediction).
The discrepancies of $R_b$ in (\ref{eq31}, \ref{eq32})
with the SM prediction\cite{hhkm}
\begin{equation}
R_b({\rm SM},\; m_t=175\;{\rm GeV})=0.2157, \label{eq33}
\end{equation}
are larger than the 3-$\sigma$ level.
This discrepancy of $R_b$ between experiments and the SM has become more
serious than it was a year ago, since the deviation (\ref{eq32}) is
larger than
that of the previous data \cite{lep94},
$R_b(1994)=0.2202\pm 0.0020$ and also because
we can no more decrease the
discrepancy by requiring a light top quark ($m_t<150$~GeV)
after the direct measurements of $m_t$ \cite{mt1,mt2}.
It is therefore natural to consider the possibility that physics
beyond SM manifests itself in the process $Z\rightarrow b\bar{b}$.

In the MSSM, the radiative corrections to the $Zb\bar{b}$ vertex
by SUSY particles and/or the extra Higgs scalars have a possibility to
explain the discrepancy in $R_b$ by increasing
$\Gamma(Z\rightarrow b\bar{b})$ over the other partial decay widths into
light quarks. There have recently appeared a lot of works
on this subject \cite{boul,abc,wells1,kimpark,sola1,rc,dabel,wells2,%
chankow2,wagner,wang,ma}.
Here we discuss the $R_b$ problem in the MSSM and show the constraints on
the SUSY particles to explain the $R_b$ data
in (\ref{eq32})\footnote{In this paper, we use the
data (\ref{eq32}) obtained by using
$R_c=0.171$. We assume here that the discrepancy of
$R_c$ in (\ref{eq31}) is caused\cite{hagiwara} by systematic uncertainty in
flavor tagging, which is less serious for $R_b$.}
for a very simple case.
We also discuss the difficulty in the MSSM explanation of the $R_b$ problem
due to exotic decays of the top quark.

The MSSM contributions to the $Zb\bar{b}$ vertex come from the loops with
(i) $t-H^+$, (ii) $\tilde{t}-\tilde{C}$, (iii) $b-(h^0,H^0,P^0)$ and
(iv) $\tilde{b}-\tilde{N}$, where $(\tilde{t},\tilde{b})$, $H^+$, $(h^0,H^0)$
and $P^0$ denote the top and bottom squarks, the charged Higgs scalar, the
light and heavy neutral Higgs scalars and the Higgs pseudoscalar,
respectively.
The magnitudes and the signs of these loop
corrections strongly depend \cite{boul,wells1,sola1} on the value of
$\tan\beta$. When $\tan\beta$ is small, typically $\tan\beta<30$,
the loops (i,ii) are dominant, due to the large
top-quark Yukawa-coupling. When $\tan\beta$ is sufficiently large,
$\tan\beta>30$, however, the loops (iii,iv) become important because of the
large bottom quark Yukawa coupling $y_b\sim m_b\tan\beta$. In both cases,
to give a correction comparable to the present discrepancy between
(\ref{eq32}) and (\ref{eq33}),
the masses of new particles in the loops
should be very light, typically lighter
than $m_Z$.

Here we discuss only the case with small $\tan\beta$,
for brevity. In this case, the contributions (iii,iv) above are negligible.
The sign of the correction to $R_b$ by the loop
(i) with the charged Higgs scalar $H^+$ is always negative and it
worsens\cite{boul,wells1} the
discrepancy. On the other hand, the loop contribution
(ii) with charginos and the scalar top $\tilde{t}$ can give
positive correction \cite{boul,wells1} to
$R_b$. The correction by the $\tilde{t}-\tilde{C}$ loop can be
comparable to the discrepancy $R_b(1995)-R_b({\rm SM})$
only if all the following conditions are satisfied:
\begin{itemize}
\item the mass of the lighter top squark $\tilde{t}_1$ and that
of the lighter chargino $\tilde{C}_1$ are small, at most
$\tilde{t}_1,\tilde{C}_1<m_Z$,
\item $\tilde{C}_1$ contains a significant higgsino component,
\item $\tilde{t}_1$ is nearly right-handed, $\tilde{t}_1\sim\tilde{t}_R$,
\item $\tan\beta$ is very small, $\tan\beta\sim 1$.
\end{itemize}
The last three conditions are necessary\cite{wells1} to
obtain a large $\tilde{t}_1-\tilde{C}_1-b$ coupling.

Here we show typical results of our numerical calculation of $R_b$
in the MSSM for very simple cases.
Fig.~3 shows $R_b(\rm MSSM)$ in the $(M_2, \mu)$-plane,
under the following conditions
\begin{eqnarray}
&& \tan\beta=1,\;\; m(\tilde{t}_1=\tilde{t}_R)=46\;{\rm GeV},\;\;
m_{H^+}\gg m_Z, \nonumber\\
&& m({\rm other\;SUSY\;particles})\gg m_Z. \label{eq34}
\end{eqnarray}
The conditions (\ref{eq34}) are chosen for maximizing the
SUSY contribution to $R_b$.
$m_t=175$~GeV, $M_1=\frac{5}{3}M_2\tan^2\theta_W$
and the LEP-I constraints on charginos and neutralinos are
also imposed.
We can see from Fig.~3 that the allowed region of $(M_2, \mu)$ to
give $R_b(\rm MSSM)>0.2190$
(consistent with the data (\ref{eq32}) at 1-$\sigma$ level)
is very narrow and it can be explored by the coming first upgrade
of LEP I (``LEP 1.5''). The allowed region of $(M_2, \mu)$ with
$R_b(\rm MSSM)>0.2175$ (consistent with the data at 95\% C.L.) is wider,
but it is completely covered by experiments at LEP 200. Therefore, if the SUSY
contribution explains the excess of $R_b$, we can produce both
$\tilde{C}_1$ and $\tilde{t}_1$ in the LEP 1.5 or LEP 200.
In the case with large $\tan\beta$, a very light $P^0$ is
necessary\cite{boul,wells1,sola1} to give large $R_b({\rm MSSM})$.

\begin{figure}
\begin{minipage}[t]{75mm}
\epsfxsize=70mm
\centerline{\epsffile{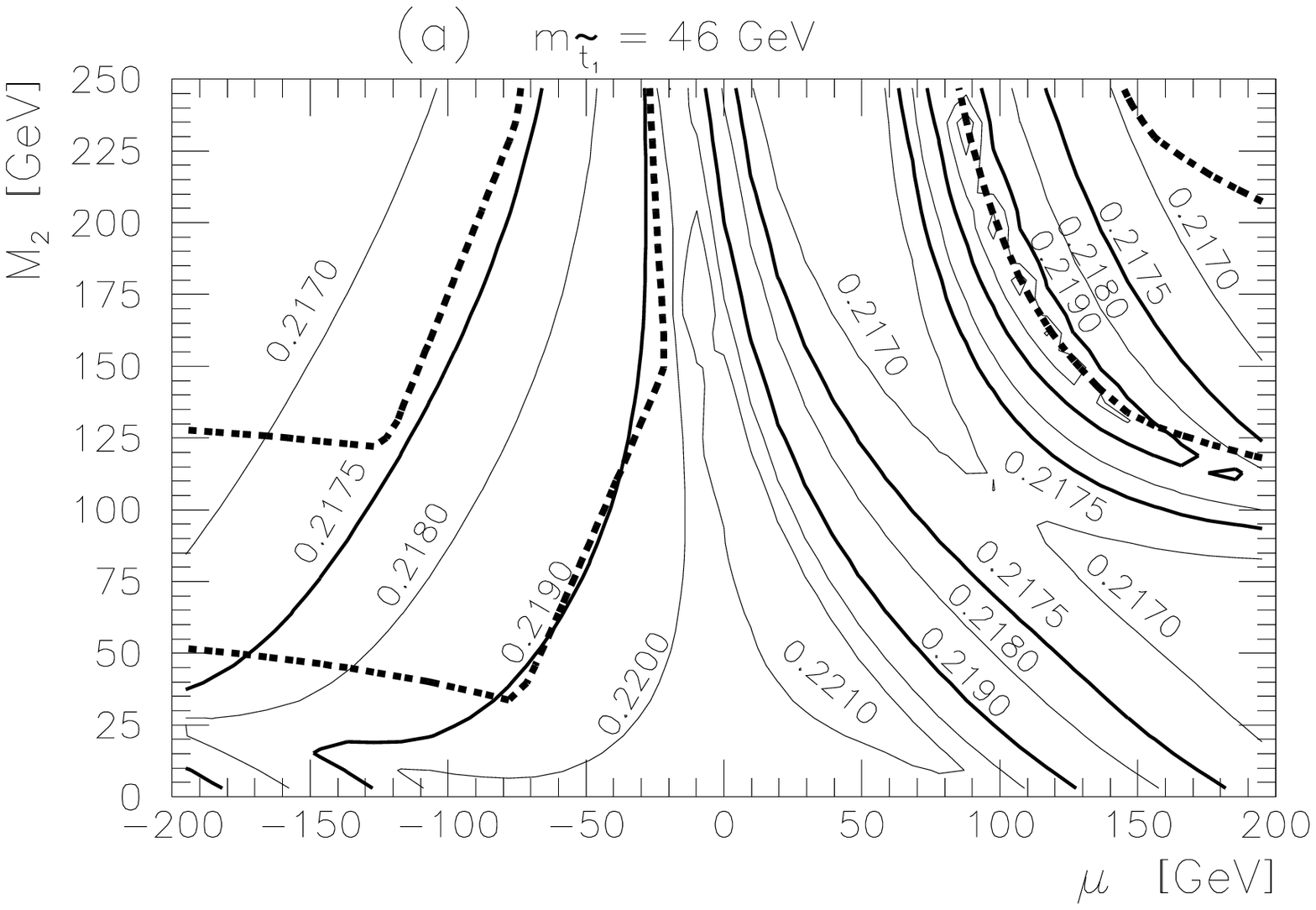}}
\end{minipage}
\begin{minipage}[t]{75mm}
\epsfxsize=70mm
\centerline{\epsffile{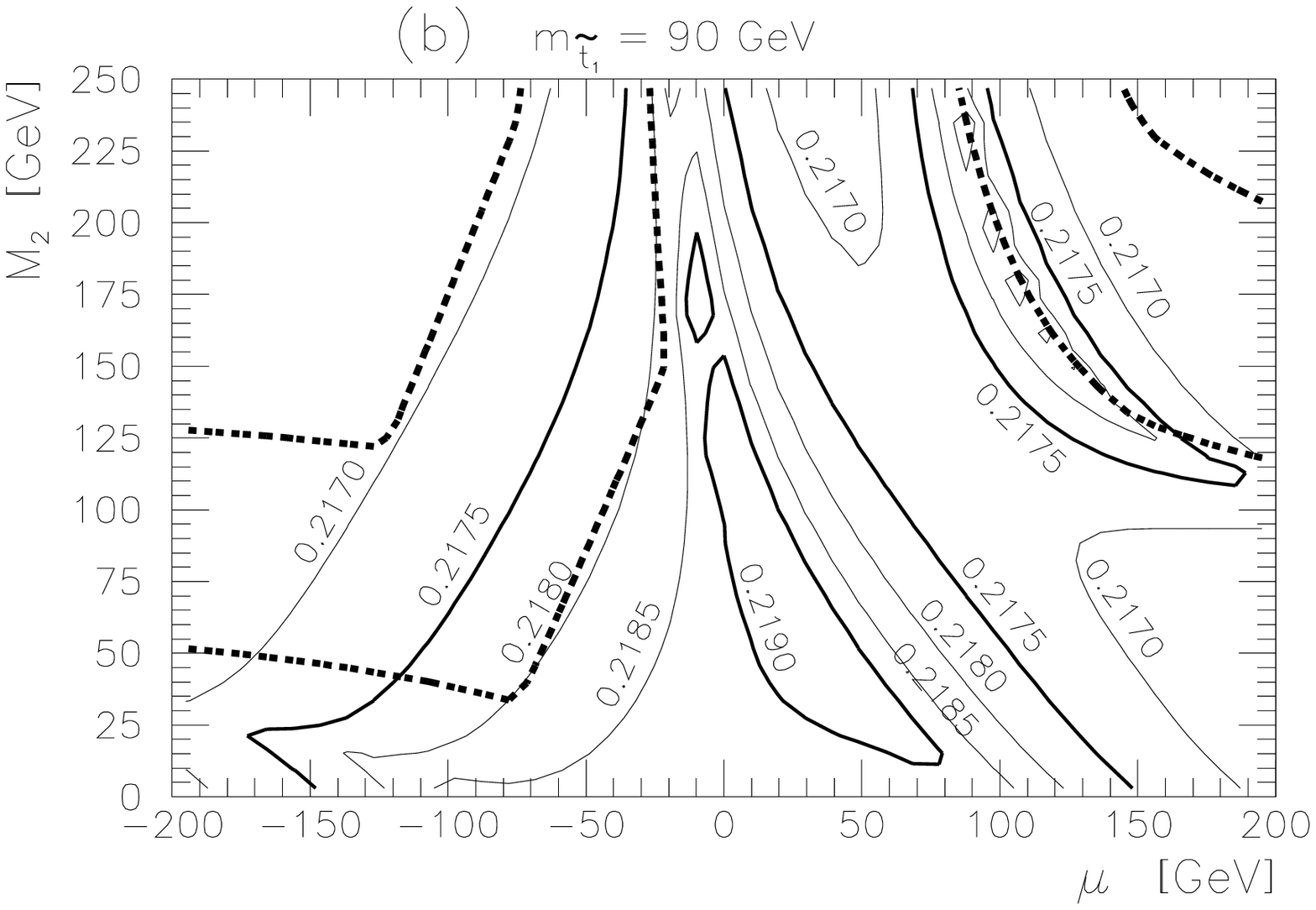}}
\end{minipage}
\caption{
$R_b$(MSSM) in the $(M_2,\mu)$ plane for $\tan\beta=1$,
$m(\tilde{t}_1=\tilde{t}_R)(\rm GeV)=$46(a), 90(b) and $m_t=175$~GeV.
The regions below the upper dashed lines can be explored by LEP 200.
The regions below the lower dashed lines are excluded by LEP-I.
}
\end{figure}

Even before the LEP upgrades, we have to check the effects of the
new particles with masses $\simgt m_Z/2$
to other experiments, which are all consistent with the SM.
In some cases, these particles cause no harm. For example,
the effect of light $\tilde{t}_R$ and $\tilde{C}_1$
to the oblique corrections ($S$, $T$) can be
sufficiently small since the SUSY contributions to $S$ and $T$
are dominated by the corrections from
left-handed sfermions \cite{chankow}.
In fact, the SUSY contribution to the $Zb\bar{b}$ vertex may
lower\cite{sola1,rc,dabel,wells2,chankow2} the
fitted value of the strong coupling constant $\alpha_s(m_Z)$, to make it
consistent with the low-energy measurements of $\alpha_s$.
However, other phenomena like the rare $b$ decay\cite{wang},
Br($b\rightarrow s\gamma$), are significantly
affected by these light new particles.

The most serious consequence of the explanation of the $R_b$ problem by light
SUSY particles is the exotic decays of the top quark \cite{wells2,wang,ma},
where exotic decay widths become comparable to the standard
mode $t\rightarrow bW^+$ which has been observed\cite{mt1,mt2}.
In the case with small $\tan\beta$, the decay mode
$t\rightarrow\tilde{t}\tilde{N}_i$ opens.
Its branching ratio is generally very large\cite{wells2,wang}.
In Fig.4, we show Br$(t\rightarrow\tilde{t}\tilde{N})$ for the
same conditions (\ref{eq34}) as in Fig.3.
As can be seen in the figure, in most of the parameter region
that gives $R_b({\rm SUSY})>0.2175$,
Br$(t\rightarrow\tilde{t}\tilde{N})$ is larger than 0.2.
In the case with large
$\tan\beta$, we instead obtain large Br$(t\rightarrow bH^+)$, which is also
exotic \cite{ma}. Although
some articles\cite{rc,mrenna} consider the possibility of
detecting these exotic decays,
the measurements of the top quark production \cite{mt1,mt2},
which suggest $\sigma_{\rm exp}(p\bar{p}\rightarrow t\rightarrow bW^+)
\ge
\sigma_{\rm SM}(p\bar{p}\rightarrow t)$ at present,
put severe upper limit on the exotic $t$-decay branching fraction
and disfavor the explanation of the $R_b$
problem by light SUSY particles.

\begin{figure}
\begin{minipage}[t]{75mm}
\epsfxsize=70mm
\centerline{\epsffile{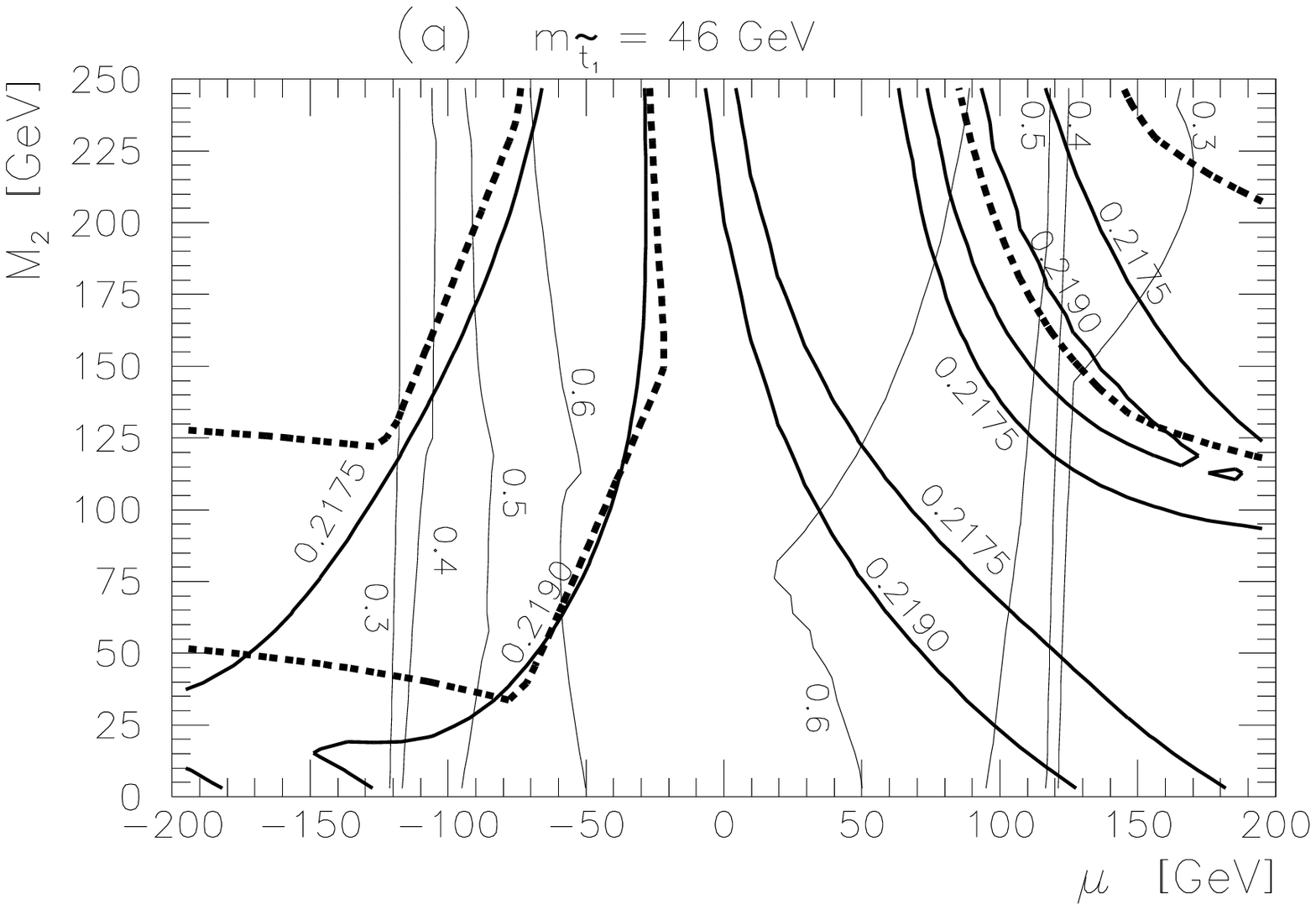}}
\end{minipage}
\begin{minipage}[t]{75mm}
\epsfxsize=70mm
\centerline{\epsffile{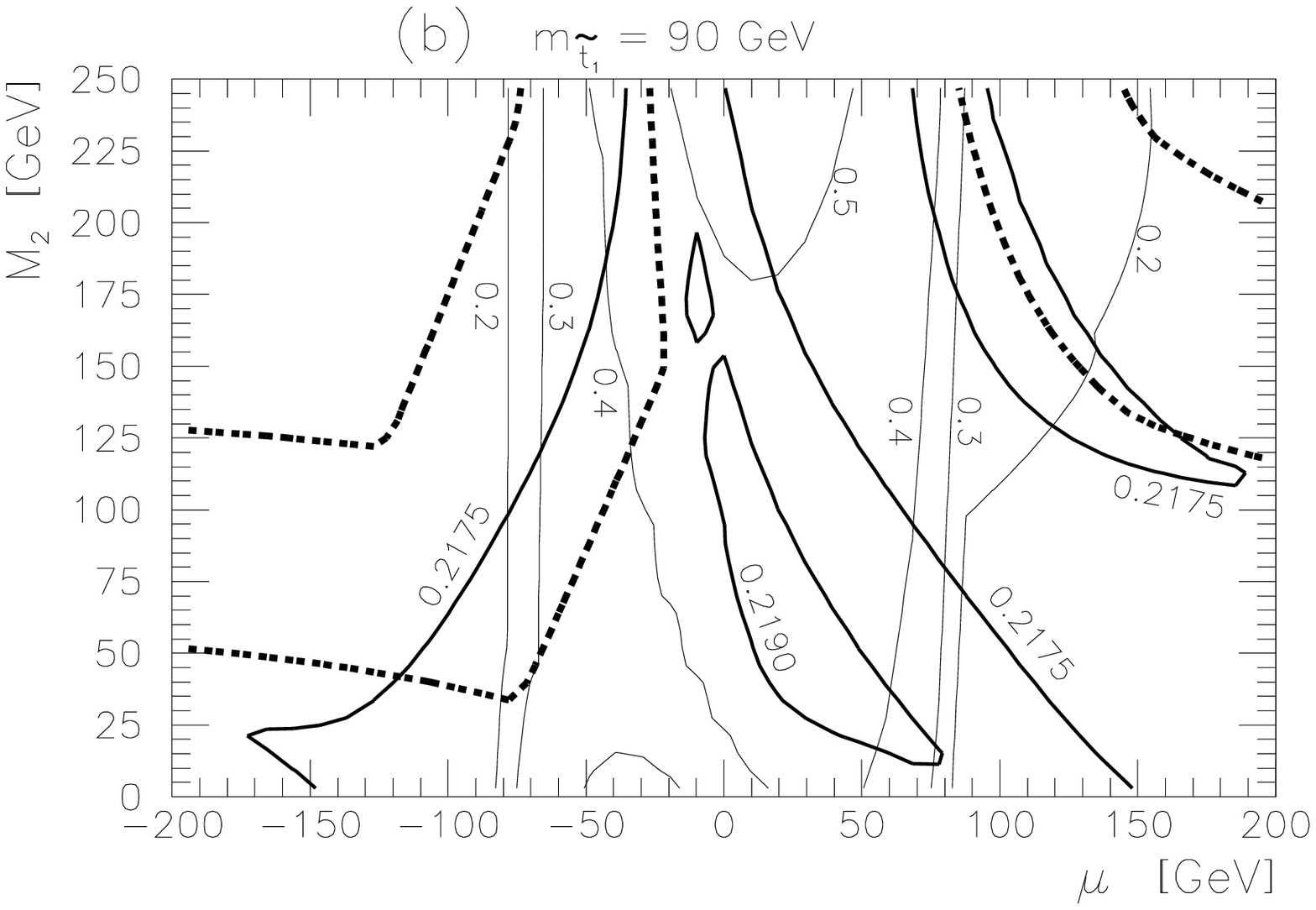}}
\end{minipage}
\caption{
The branching ratio Br($t\rightarrow\tilde{t}\tilde{N}$) in the
$(M_2,\mu)$ plane.
The experimental bounds by LEP-I (lower dashed lines) and
LEP 200 (upper dashed lines),
and the region with $R_b(\rm MSSM)>$0.2190, 0.2175 (thick solid lines)
are also shown.
Parameters are the same as in Fig.3.
}
\end{figure}

\section{Summary}
We have discussed the SUSY contributions to the quark-lepton universality
violation in charged currents and the anomaly in $R_b$ at the $Z$ pole.
In both cases, the non-oblique radiative corrections play crucial role.
We have studied the constraints on the SUSY particles to explain
recent measurements of above two observables, which suggest physics beyond
the Standard Model.
The observed 1-$\sigma$ violation of the quark-lepton universality,
if it turns out to be real, may be explained by relatively light
sleptons and chargino. In contract, although it is possible to
explain the 3-$\sigma$ excess of $R_b$ by very light scalar top and
chargino, this possibility is severely restricted by the
upper limit for exotic decays of the top quark.

{\it Note added.} After the seminar, we received a paper\cite{wells3}
in which the analysis using the recent data (\ref{eq32}) was given.
Ref.\citen{wells3} claims that the SUSY explanation with high
$\tan\beta$ and light $P^0$ is ruled out by constraints from
decays $Z\rightarrow b\bar{b}P^0$ and $b\rightarrow c\tau\bar{\nu}_{\tau}$.

\section*{Acknowledgements}
We thank V.~Barger for careful reading of the manuscript.
The work of Y.~Y. is supported in part by the JSPS Fellowships
and the Grant-in-Aid for Scientific Research from the Ministry
of Education, Science and Culture of Japan No. 07-1923.

\end{document}